\documentclass[letter,scriptaddress,twocolumn, prl,showpacs,superscriptaddress,showkeys,groupedaddress]{revtex4}

       \usepackage{amsmath}
       \usepackage{makeidx}
       \usepackage{amsfonts}
       \usepackage[ansinew]{inputenc}
       \usepackage[usenames,dvipsnames]{pstricks}
       \usepackage{subfigure}
       \usepackage{epsfig}
       \usepackage{pst-grad} 
       \usepackage{pst-plot} 
       \usepackage[colorlinks,hyperindex]{hyperref}

\usepackage{epsfig,color}
\usepackage{pdfsync}

\usepackage{epsfig}
\usepackage{lscape}
\usepackage{amssymb}
\usepackage{graphicx}
\usepackage{graphics}

 \def\be{\begin{equation}}
 \def\ee{\end{equation}}
 \def\bea{\begin{eqnarray}}
 \def\eea{\end{eqnarray}}
 \def\bean{\begin{eqnarray*}}
 \def\eean{\end{eqnarray*}}
 \def\gsim{\mathrel{\rlap{\lower0.2em\hbox{$\sim$}}\raise0.2em\hbox{$>$}}}
 \def\ksim{\mathrel{\rlap{\lower0.2em\hbox{$\sim$}}\raise0.2em\hbox{$<$}}}
 \def\kg{\mathrel{\rlap{\lower0.25em\hbox{$>$}}\raise0.25em\hbox{$<$}}}

\newcommand{\bs}[1]{\boldsymbol #1}
\newcommand{\bsp}[1]{p\!\!\!\!\! p \!\!\!\!\! p\!\!\!\!\! p}
\newcommand{\bsq}[1]{q\!\!\!\!\! q}
\newcommand{\bsA}[1]{A\!\!\!\!\! A}
\newcommand{\bsP}[1]{P\!\!\!\!\! P}
\newcommand{\bsAA}[1]{A\!\!\!\!\!\!\! A}
\newcommand{\bsa}[1]{a\!\!\!\!\! a}
\newcommand{\bsu}[1]{u\!\!\!\!\! u}
\newcommand{\bsbt}[1]{\beta\!\!\!\!\!\! \beta \!\!\!\!\!\! \beta \!\!\!\!\!\! \beta}

\begin{document}

\title{Transport coefficients of heavy quarks around $T_c$ at finite quark chemical potential}

\author{H.~Berrehrah}
\affiliation{\begin{minipage}[c]{0.9\textwidth}Frankfurt Institute for Advanced Studies, Johann Wolfgang Goethe Universit\"at, Ruth-Moufang-Strasse 1,60438 Frankfurt am Main, Germany\end{minipage}}

\author{P.B.~Gossiaux}
\affiliation{\begin{minipage}[c]{0.9\textwidth}Subatech, UMR 6457, IN2P3/CNRS, Universit\'e de Nantes, \'Ecole des Mines de Nantes, 4 rue Alfred Kastler, 44307 Nantes cedex 3, France\end{minipage}}

\author{J.~Aichelin}
\affiliation{\begin{minipage}[c]{0.9\textwidth}Subatech, UMR 6457, IN2P3/CNRS, Universit\'e de Nantes, \'Ecole des Mines de Nantes, 4 rue Alfred Kastler, 44307 Nantes cedex 3, France\end{minipage}}

\author{W.~Cassing}
\affiliation{\begin{minipage}[c]{0.9\textwidth}Institut f\"ur Theoretische Physik, Universit\"at Giessen, 35392 Giessen, Germany\end{minipage}}

\author{J.M. Torres-Rincon}
\affiliation{\begin{minipage}[c]{0.9\textwidth}Subatech, UMR 6457, IN2P3/CNRS, Universit\'e de Nantes, \'Ecole des Mines de Nantes, 4 rue Alfred Kastler, 44307 Nantes cedex 3, France\end{minipage}}

\author{E.~Bratkovskaya}
\affiliation{\begin{minipage}[c]{0.9\textwidth}Frankfurt Institute for Advanced Studies, Johann Wolfgang Goethe Universit\"at, Ruth-Moufang-Strasse 1,60438 Frankfurt am Main, Germany\end{minipage}}

\begin{abstract}
The interactions of heavy quarks with the partonic environment at finite temperature $T$ and finite quark chemical potential $\mu_q$ are investigated in terms of transport coefficients within the Dynamical Quasi-Particle model (DQPM) designed to reproduce the lattice-QCD results (including the partonic equation of state) in thermodynamic equilibrium. These results are
confronted with those of nuclear many-body calculations close to the critical temperature $T_c$.  The hadronic and partonic
spatial diffusion coefficients join smoothly and show a pronounced minimum around $T_c$, at $\mu_q=0$ as well as at finite  $\mu_q$. Close and above $T_c$ its absolute value matches the lQCD calculations for $\mu_q=0$.
The smooth transition of the heavy quark transport coefficients from the hadronic to the partonic medium
 corresponds to a cross over  in line with  lattice calculations, and differs substantially from perturbative QCD (pQCD) calculations which show a large discontinuity at $T_c$. This indicates that in the vicinity of $T_c$ dynamically dressed massive partons and not massless pQCD partons are the effective degrees-of-freedom in the quark-gluon plasma.
\end{abstract}

\maketitle

The primary goal of ultra-relativistic heavy-ion experiments at the Relativistic Heavy-Ion Collider (RHIC) and the Large Hadron Collider (LHC) is to search and to characterize a new state of matter, the Quark-Gluon-Plasma (QGP),  composed of partons as predicted by lattice calculations of Quantum Chromo Dynamics (lQCD). Heavy quarks are considered to be a suitable tool to study the properties of the QGP since they provide an independent scale by their heavy mass which is large compared to the temperatures that are achieved at RHIC or LHC energies in heavy-ion collisions. Initially, the heavy quarks are produced in primary hard $NN$ collisions, which can be calculated in pQCD \cite{Cacciari:2005rk} and controlled by $pp$ experimental data \cite{Adamczyk:2012af}. The modification of the heavy quark properties (spectral functions) in a QGP is expected to be small and their interaction is sensitive to the time evolution of the QGP essentially consisting of light quarks/antiquarks and gluons. Therefore, the difference between the initial heavy quark spectrum at production and that of final $D$ and $B$ mesons, observed asymptotically in the detector, is an image of the properties of the QGP during its expansion.

Different models \cite{Gossiaux:2008jv,Gossiaux:2010yx,Gossiaux:2009mk,Moore:2004tg,van Hees:2005wb, Djordjevic:2007at,Cao:2013ita,Hamza1} have been advanced to describe the heavy-quark interactions in heavy-ion collisions. The most straightforward way to compare these models is via transport coefficients which condense the complicated interactions to a few characteristic quantities. In this Letter we study the transport coefficients at finite $T$ and finite quark chemical potential $\mu_q$ based on the DQPM model \cite{Cassing:2009vt,Cassing:2008nn,Cassing:2007yg}. In the calculations presented here we limit ourselves to Born type matrix elements. The calculations using the pole masses of the parton spectral functions and coupling constants of the DQPM model are therefore called Dressed pQCD (DpQCD) calculations.
We compare our results with nuclear many-body calculations below $T_c$ \cite{Torres-Rincon:2013nfa} and with pQCD  and lQCD results at finite $T > T_c$  \cite{Moore:2004tg,Banerjee:2011ra}.

The DQPM describes QCD properties in terms of 'resumed' single-particle Green's functions (in the sense of a two-particle irreducible (2PI) approach). In other words: the degrees-of-freedom of the QGP can be interpreted as being strongly interacting massive effective quasi-particles with broad spectral functions (due to the high interaction rates). The dynamical quasiparticle entropy density $s^{DQP}$ has been fitted to lQCD data which allows to fix for $\mu_q=0$ the 3 parameters of the DQPM entirely (cf. Refs. \cite{Cassing:2009vt,Cassing:2008nn,Cassing:2007yg} for the details of the DQPM model).

The DQPM employs a Lorentzian parametrization of the partonic spectral functions $A_{i} (\omega_{i})$,
where $i$ is the parton species:
{
\begin{eqnarray}
\label{equ:Sec2.2}
A_i (\omega_i) & & \ = \frac{\gamma_i}{\tilde{E}_i} \biggl(\frac{1}{(\omega_i - \tilde{E}_i)^2 + \gamma_i^2} -
\frac{1}{(\omega_i + \tilde{E}_i)^2 + \gamma_i^2} \biggr)
\nonumber\\
& & {} \equiv \frac{4 \omega_i \gamma_i}{(\omega_i^2 - \bs{p}_i^2 - M_i^2)^2 +
4 \gamma_i^2 \omega_i^2},
\end{eqnarray}}
with $\tilde{E}_i^2 (\bs{p}_i) = \bs{p}_i^2 + M_i^2 - \gamma_i^2$, and $i \in [g, q, \bar{q}, Q, \bar{Q}]$.
The spectral functions $A_i (\omega_i)$ are normalized as:
{\setlength\arraycolsep{0pt}
\begin{eqnarray}
\int_{- \infty}^{+ \infty} \frac{d \omega_i}{2 \pi} \ \omega_i
\ A_i (\omega_i, \bs{p}) = \!\!\! \int_0^{+ \infty} \frac{d \omega_i}{2 \pi} \ 2 \omega_i \ A_i (\omega_i, \bs{p}_i) =
1 , \nonumber
\end{eqnarray}}
where $M_i$, $\gamma_i$ are the dynamical quasi-particle mass (i.e. pole mass) and width of the
spectral function for particle $i$, respectively. They are directly related to the real and imaginary parts of the related self-energy, e.g. $\Pi_i = M_i^2 - 2 i \gamma_i \omega_i$, \cite{Cassing:2009vt}. In the off-shell approach, $\omega_i$ is an independent variable and related to the \emph{``running mass''} $m_i$ by: $\omega_i^2 = m_i^2 + \bs{p}_i^2$. The mass (for gluons and quarks) is assumed to be given by the thermal mass in the asymptotic high-momentum regime. We note that this approach is consistent with respect to microcausality in field theory \cite{Rauber}.

The DQPM model has originally been designed to reproduce the QCD equation of state, calculated on the lattice, at zero chemical potential $\mu_q$ in an effective quasiparticle approach. Since the EoS from lattice QCD is still rather unknown at finite $\mu_q$, we use the assumption that the  energy density at $T_c(\mu_q)$ is approximately  independent of the chemical potential $\mu_q$ which provides $T_c(\mu_q)$ implicitly. Assuming, furthermore, that the chemical potential is identical for all flavors and employing the functional form of the parton masses from hard thermal loop calculations \cite{Vija:1994is,Peshier:1999ww,Ozvenchuk:2012kh}:
\begin{eqnarray}
\label{equ:Sec2.6}
& & M_g^2 (T) = \frac{g^2 (T/T_c)}{6} \Biggl((N_c + \frac{1}{2} N_f) T^2 + \frac{N_c}{2}
\sum_q \frac{\mu_q^2}{\pi^2} \Biggr) \ ,
\nonumber\\
& & {} M_q^2 (T) = \frac{N_c^2 - 1}{8 N_c} g^2 (T/T_c) \Biggl( T^2 + \frac{\mu_q^2}{\pi^2}
\Biggr)\ ,
\end{eqnarray}
we find that a finite chemical potential $\mu_q$ can be accommodated by introducing an effective temperature
\begin{equation}
\label{equ:Sec2.7} T^{*2} = T^2+\frac{\mu_q^2}{\pi^2}.
\end{equation}
The extrapolation  to finite $\mu_q$ within the DQPM --
assuming a constant energy density at $T_c(\mu_q)$ -- leads to the
approximation
\begin{equation}
\label{equ:Sec2.8} T_c(\mu_q) = T_c(\mu_q=0) \ \sqrt{1 - d^2 \mu_q^2}
\end{equation}
with $d^2= 8.79$ GeV$^{-2}$ and $T_c(\mu_q=0) = 158$ MeV. Furthermore,  replacing in the width $\gamma(T/T_c)$ and in the coupling constant $g(T/T_c)$ the dimensionless quantity $T/T_c$ by $h=T^*/T_c(\mu_q)$ we have 
\begin{eqnarray}
\label{equ:Sec2.9}
&&\hspace*{-0.78cm}\gamma_g (T,\mu)\!=\!\frac{1}{3} N_c \frac{g^2 (h)}{8 \pi} \, T \ln \!\left(\!\frac{2 c}{g^2(h)}+1\! \right)
\nonumber\\
&&\hspace*{-0.78cm}\gamma_q (T,\mu)\!=\!\frac{1}{3}\frac{N_c^2 - 1}{2 N_c}\frac{g^2 (h)}{8 \pi}\,T \ln \!\left(\!\frac{2 c}{g^2 (h)}+1 \!\right)\nonumber\\
&&\hspace*{-0.78cm}\displaystyle{g^2( T^*/T_c(\mu))\!=\!\!\frac{48 \pi^2}{(11 N_c - 2 N_f) \ln\!\left(\!\lambda^2 (h - \frac{T_s}{T_c(\mu_q)})^2\!\right)}}
\end{eqnarray}
with $\lambda = 2.42$, c=14.4  and $T_s = 73$  MeV. Eqs. (\ref{equ:Sec2.8}) and (\ref{equ:Sec2.9}) define the DQPM ingredients necessary for the calculations at finite temperature and quark chemical potential $\mu_q$. In this Letter we limit ourselves to on-shell quarks and gluons because we have found in Ref.  \cite{Berrehrah:2014kba} that a finite width $\gamma_{g,q,Q}$ for gluons $g$, light $q$ and heavy quarks $Q$ has an impact of about 10-20\% on the heavy-quark  transport coefficients.

Having  the masses and the coupling constant specified  we can calculate transport coefficients  $\mathcal{X}$ defined by \cite{Berrehrah:2014kba}
\begin{eqnarray}
\label{equ:Sec1.1}
&&\hspace*{-0.7cm}\frac{d <\!\!\mathcal{X}\!\!>}{d t} = \sum_{q,g} \frac{1}{(2 \pi)^5 2 E_Q} \int \frac{d^3 q}{2 E_q} f (\bs{q})
\int \frac{d^3 q'}{2 E_{q'}} \int \frac{d^3 p'_Q}{2 E'_{Q}}\nonumber \\
& & \hspace*{1.2cm} \times \ \delta^{(4)} (P_{in} - P_{fin}) \ \mathcal{X} \ {\frac{1}{g_Q g_p}}
|\mathcal{M}_{2,2}|^2\!,
\end{eqnarray}
where $p'_Q$ ($ E'_Q$) is the final momentum (energy) of the heavy quark with the initial energy  $E_Q$. $q$ ($E_q$) and $q'$ ($ E_{q'}$) are the initial and final momenta (energies) of the massless partons and $f(\bs{q})$  is their thermal distribution whereas $|\mathcal{M}_{2,2}|^2$  is the transition matrix-element squared for 2 $\to$ 2 scattering . In (\ref{equ:Sec1.1}) $g_Q$ is the degeneracy factor of the heavy quark ($g_Q = 6$) and $g_p$ is the degeneracy factor of the massless partons ($g_p$=16 for gluons and $g_p$=6 for light quarks). Employing $\mathcal{X}$ = $(E - E')$  we can calculate the energy loss, $\frac{d< E>}{d t} (p_Q, T)$, whereas  $\mathcal{X}$ = $(\bs{p}_Q - \bs{p}'_Q)$ gives the drag  coefficient,$\frac{d <\bs{p}_Q>}{d t}  = A (p_Q, T)$.

The spatial diffusion coefficient $D_s$  can be expressed in two different ways \cite{Moore:2004tg}.
It can be obtained from the  slope of the drag coefficient divided by the heavy quark
momentum $\eta_D = A/p_Q$,
\begin{equation}
D_s =\lim_{p_Q\to 0} T/(M_Q \eta_D),
\label{eq7} \end{equation}
as  in \cite{Berrehrah:2014kba}. It can also be obtained from the diffusion coefficient
$\kappa = \frac{1}{3}\frac{d<(\bs{p}_Q - \bs{p}'_Q)^2>}{ dt}$,
calculated with eq.  (\ref{equ:Sec1.1}), as  \cite{Torres-Rincon:2013nfa}
\begin{equation}
D_s = \lim_{p_Q\to 0} \frac{\kappa}{2 M_Q^2 \eta_D^2} .
\label{eq9} \end{equation}
Both definitions agree if  the Einstein relation is valid. In the case of the DpQCD  model
the deviation from the Einstein relation for small
momenta $p_Q$ is of the order 10-15\%. We will adopt   eq.(\ref{eq7}) for the
calculations in this Letter.
\begin{figure} 
\begin{center}
\includegraphics[scale=.48]{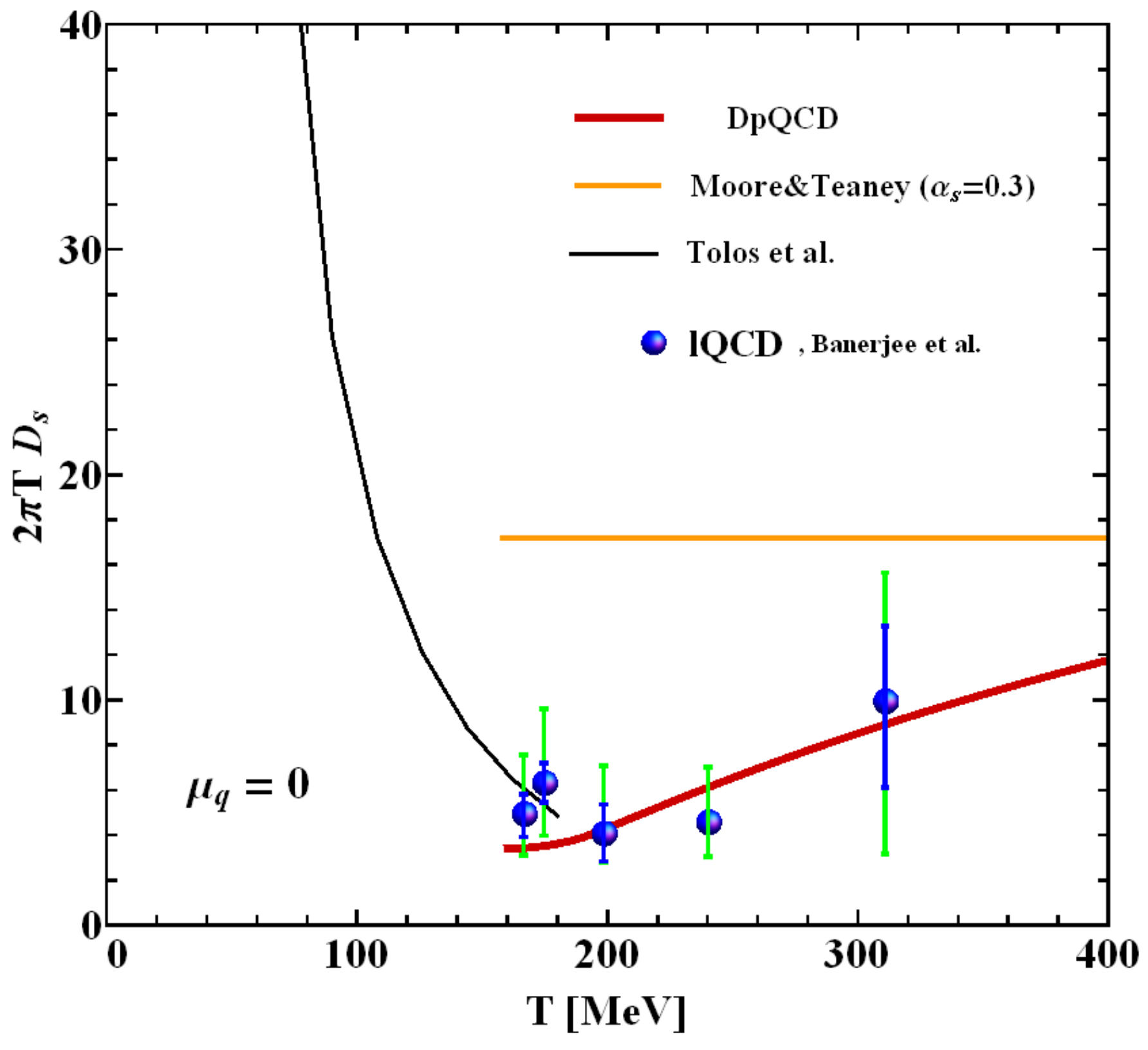}
\caption{\emph{(Color online)  Spatial diffusion coefficient for heavy quarks, $D_s$, as a function of
 $T$ for $\mu_q=0$.  Below $T=180\  MeV$ we display the hadronic diffusion
 coefficient  from
 \cite{Torres-Rincon:2013nfa}, above $T=180$ MeV that for a partonic environment. The solid orange
 line is the result of \cite{Moore:2004tg} while the red thick solid line shows the DpQCD prediction.
 The lattice calculations are from Ref. \cite{Banerjee:2011ra}.}}
\label{fig:cdEdxTmu}
\end{center}
\vskip - 1cm
\end{figure}

The relation (\ref{eq7}) is strictly valid in the non-relativistic limit where bremsstrahlung is negligible,
i.e. for velocities
$\gamma \ v < 1/\sqrt{\alpha_s}$ to leading logarithm in $T/m_D$, where $m_D$ is the Debye mass.
Therefore, it is a good approximation for the interaction of thermal heavy quarks, $M_Q \gg T$, with a
typical thermal momentum $p \sim \sqrt{M T}$ and a velocity $v \sim \sqrt{T/M} \ll 1$.

 In Fig.\ref{fig:cdEdxTmu} we display $D_s$ (\ref{eq9}) as a function of $T$ for $\mu_q = 0$. Our results are compared with the results obtained by Moore and Teaney \cite{Moore:2004tg} for massless partons and $\alpha_s $ = 0.3 as well as with the lattice calculations from Ref. \cite{Banerjee:2011ra}. These lattice results  have recently been confirmed by the Bielefeld collaboration  \cite{kaz}. The spatial diffusion coefficient in deconfined matter is compared with the result for the spatial diffusion coefficient of a heavy meson in hadronic matter  \cite{Torres-Rincon:2013nfa}.
 We observe that at $T=T_c$ the spatial diffusion coefficients for hadronic and partonic matter join almost continuously and agree with the lattice results. On the other hand, pQCD calculations yield a large value of the spatial diffusion coefficient as compared to the DpQCD model leading to a discontinuity of $D_s$ close to $T_c$.  Rapp et al. \cite{Rapp:2009my} have shown that the spatial diffusion coefficient in pQCD calculations can be lowered by adding  nonperturbative heavy-quark interactions. Also hard thermal loop calculations
with effective Debye masses and a running coupling constant lead to a substantial lowering
and bring $D_s$ to the vicinity of the lattice results  \cite{Berrehrah:2014kba}.
\begin{figure} 
\begin{center}
\includegraphics[height=8cm, width=8.5cm]{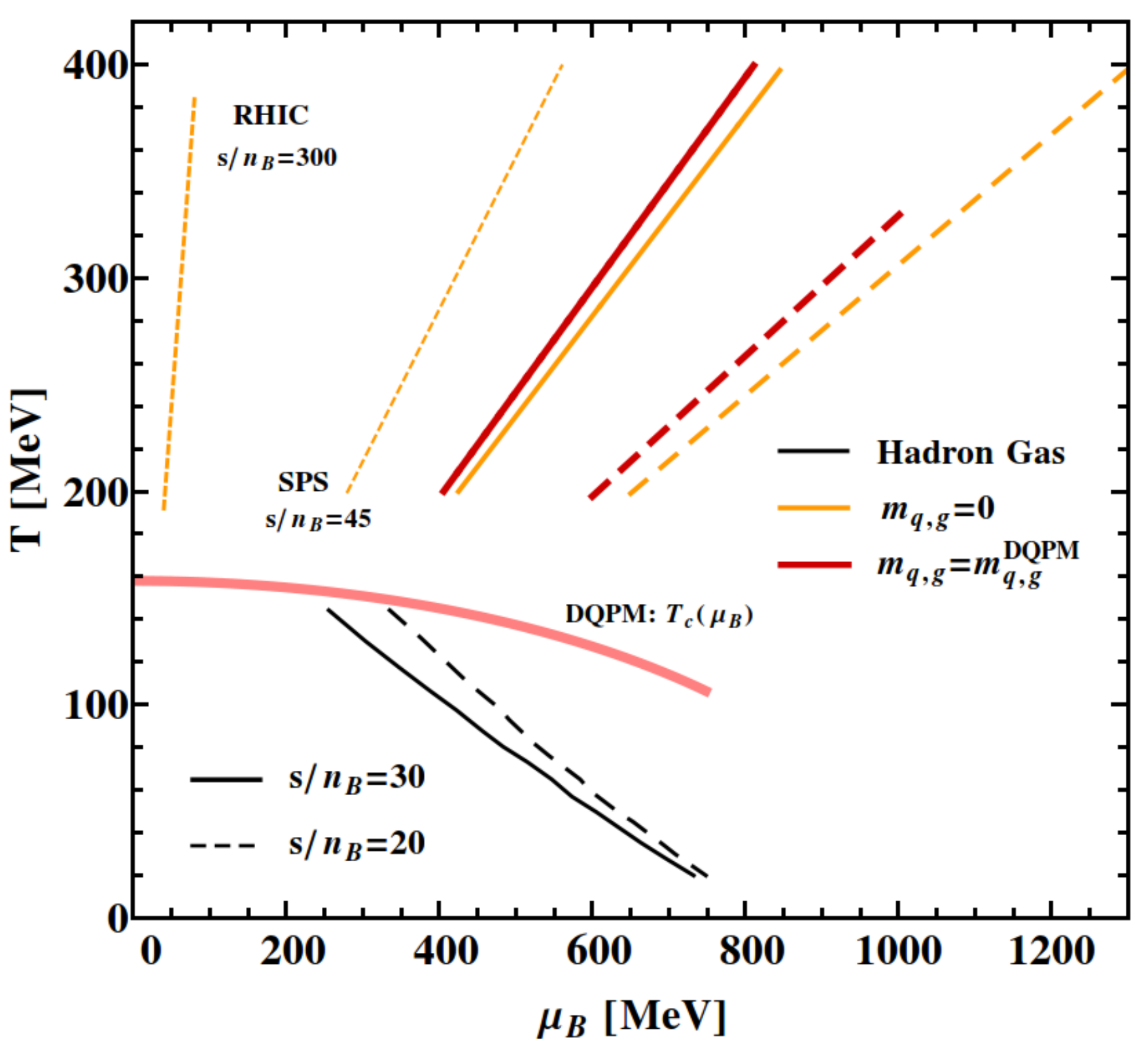}
\caption{\emph{(Color online) Adiabatic trajectories in the plasma and the hadron phase as a
function of $T$ and $\mu_B = 3\mu_q$. For the plasma phase we present calculations in the DpQCD
model
as well as for massless partons. The trajectories in the hadron phase are taken from  \cite{Torres-Rincon:2013nfa}.
 We display as well the transition temperature $T_c(\mu_B)$ between the hadronic and plasma phase as given
 in the DQPM model.}}
\label{TvsmuAdiabaticFAIR}
\end{center}
\vskip - 1cm
\end{figure}

\begin{figure} 
\begin{center}
\includegraphics[scale=.48]{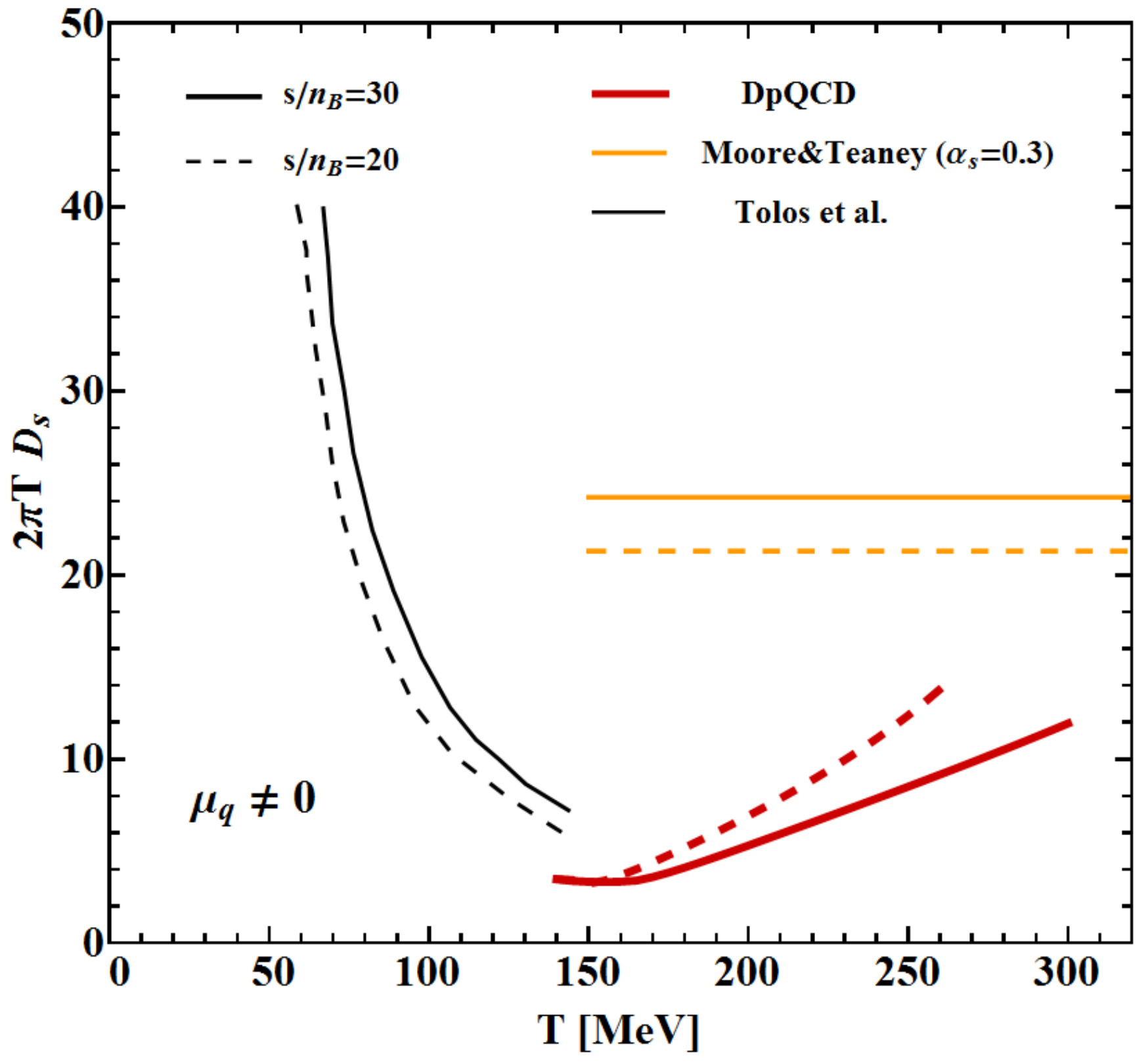}
\caption{\emph{(Color online) Spatial diffusion constant, $D_s$, as a function of $T$ for $\mu_q \neq 0$.
$D_s$ is displayed for different values of $s/n_B$ for a hadronic environment \cite{Torres-Rincon:2013nfa}
as well as for a partonic environment. For the latter pQCD calculations are confronted with DpQCD
calculations. }}
\label{mune0}
\end{center}
\vskip - 0.44cm
\end{figure}

These calculations can be extended to finite $\mu_q$ assuming adiabatic trajectories for the expansion.
In Fig. \ref{TvsmuAdiabaticFAIR} we display trajectories of constant entropy  per net baryon for the
hadronic as well as for the partonic phase. The latter is calculated using the effective masses of the
plasma constituents in the DQPM model as well as for massless partons. For a given $s/n_B$ the chemical
 potential $\mu_B$ is a monotonic function of $T$ and therefore we can display  $D_s$ as a function
 of $T$  and  $s/n_B$. Fig. \ref{mune0} displays the spatial diffusion coefficient for finite chemical
 potential, i.e for different values of the entropy per net baryon $s/n_B$. The pQCD calculations are
 obtained by adding the chemical potential to the thermal distributions and to the Debye mass when
 calculating the pQCD drag and diffusion coefficients (cf. Eq.(B13) of Ref.\cite{Moore:2004tg}).
 We observe - as in the $\mu_q = 0$ case - that the DpQCD spatial diffusion coefficient of
 heavy quarks approximately joins smoothly those of the hadron gas. (We expect that in the $\mu_B$
 region investigated here
the transition remains a cross over transition). On the contrary, pQCD
 calculations close to $T_c$ are a factor of 3 higher leading to a discontinuity of the spatial
 diffusion coefficient, not compatible with a cross over transition as predicted by lattice calculations.
 This is a strong indication that close to the phase transition the effective degrees-of-freedom are massive
 quasi-particles and not massless quarks and gluons.

\begin{figure}[h!] 
\begin{center}
\includegraphics[scale=.41]{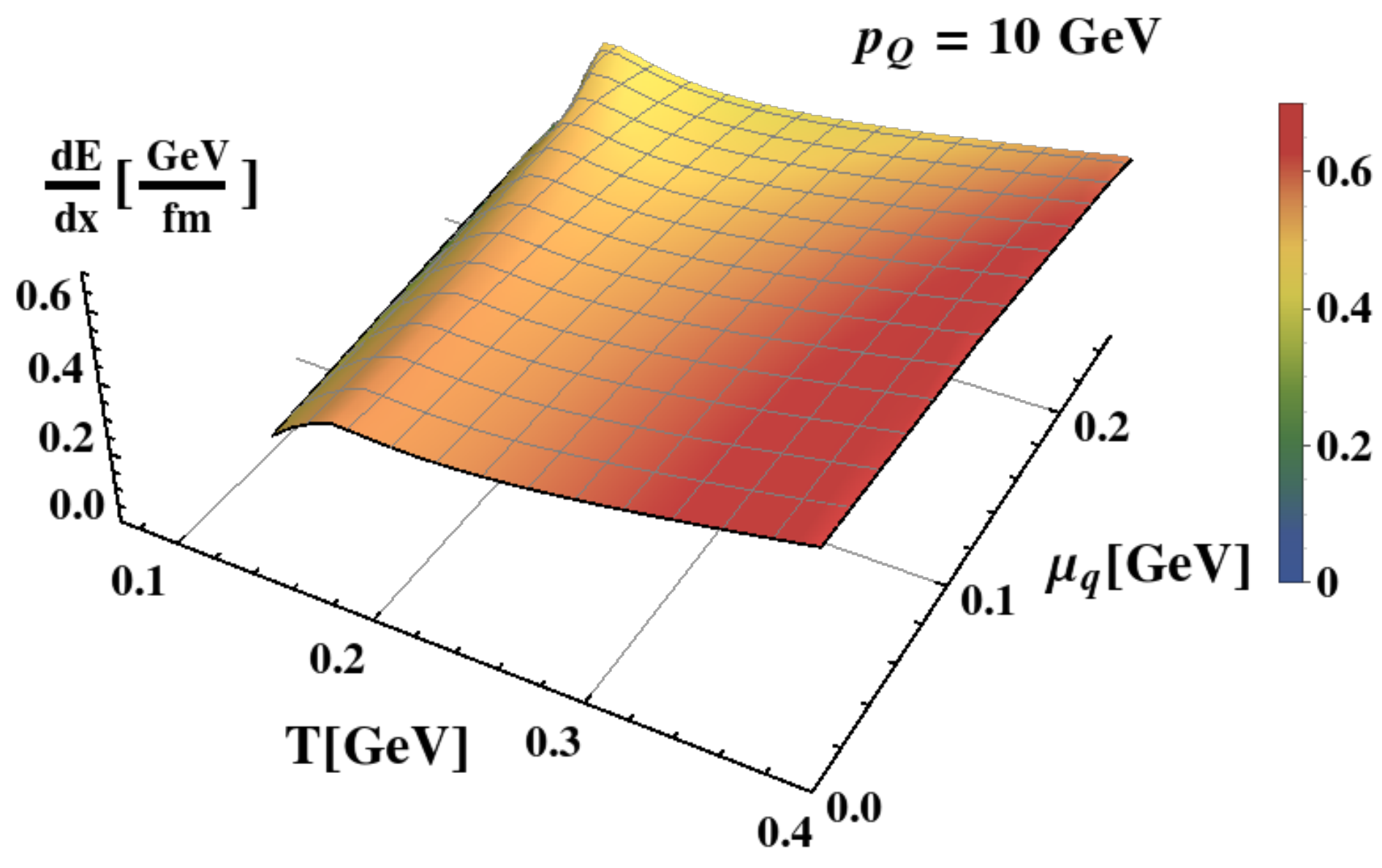}
\caption{\emph{(Color online) Energy loss per unit length, $dE/dx$, of a c - quark with incoming momentum
of $10\ GeV$ in the plasma rest frame as a function of the temperature and quark chemical potential.}}
\label{fig:EdxTmu}
\end{center}
\vskip - 0.1cm
\end{figure}

For comparing our model predictions with experimental data another transport coefficient, the energy loss of
a heavy quark per unit length, $\frac{d<E>}{dx}=  \frac{d<E>}{vdt}$, is important. It can be obtained
from eq. (\ref{equ:Sec1.1}) by the choice $\mathcal{X}$ = $(E_Q - E'_Q)$. The energy loss  of a heavy quark
with an incoming momentum of  10 GeV/c as a function of $T$ and $\mu_q$  in the DpQCD approach is presented
in Fig. \ref{fig:EdxTmu}. As expected for a cross-over transition we observe a very smooth dependence on
both variables, $T$ and $\mu_q$. For $\mu_q =0$ the gluon mass depends on the temperature and
therefore the
increase of the energy loss is due to the change of the coupling constant.
For $\mu_q =0.2$  GeV, the energy loss is also increasing with temperature but less than for $\mu_q=0 $
because here the increase of the coupling constant is partially counterbalanced by the decrease of
the effective gluon mass.

We recall that the properties of the QCD medium in terms of the shear viscosity over entropy ratio $\eta/s$ as
well as the electric conductivity over temperature $\sigma/T$ show a minimum close to $T_c$
(cf. \cite{Marty:2013ita,Cassing:2013iz}) which apparently repeats in the charm spatial diffusion coefficient
reflecting a maximum in the interaction strength of the QCD degrees-of-freedom at temperatures close to $T_c$.

\vspace*{2mm}
We acknowledge  the ``HIC for FAIR'' framework of the ``LOEWE'' program , the European program I3- Hadron
Physics and the program "Together" of the region Pays de la Loire for support of this
work and valuable discussions with V. Ozvenchuck, R. Marty  and T. Steinert.


\end{document}